Article

# Comparison of Topic Modelling Approaches in the Banking Context

Bayode Ogunleye [1,*], Tonderai Maswera [1], Laurence Hirsch [1], Jotham Gaudoin [1] and Teresa Brunsdon [2]

1 Department of Computing, Sheffield Hallam University, Sheffield S1 2NU, UK
2 Department of Statistics, University of Warwick, Coventry CV4 7AL, UK
* Correspondence: b.ogunleye@shu.ac.uk

**Abstract:** Topic modelling is a prominent task for automatic topic extraction in many applications such as sentiment analysis and recommendation systems. The approach is vital for service industries to monitor their customer discussions. The use of traditional approaches such as Latent Dirichlet Allocation (LDA) for topic discovery has shown great performances, however, they are not consistent in their results as these approaches suffer from data sparseness and inability to model the word order in a document. Thus, this study presents the use of Kernel Principal Component Analysis (KernelPCA) and K-means Clustering in the BERTopic architecture. We have prepared a new dataset using tweets from customers of Nigerian banks and we use this to compare the topic modelling approaches. Our findings showed KernelPCA and K-means in the BERTopic architecture-produced coherent topics with a coherence score of 0.8463.

## 1. Introduction

Due to the increasing availability of big data, the need for topic modelling (TM) approaches for topic discovery is increasing [1]. For example, service industries such as banks sell different products and services. Thus, they need the main topics of their customers' enquiries, feedback, reviews, and discussions to help them target their resources. Blei et al. [2] identified topic modelling as a type of statistical learning approach for detecting coherent topics in a document. Topic modelling (TM) is considered an important task for many applications such as aspect extraction in sentiment analysis [3–5], topic extraction for user preference in a recommender system [6], document summarization [7], topic discovery in a chat box system [8], and topic extraction for fake news detection [9]. TM has been applied successfully in different domains such as health [10], online learning platforms [11], software engineering [12], and legal documents [13] to discover hidden topics.

With the increasing use of supervised learning algorithms to provide automated systems, TM provides an opportunity to apply unsupervised learning algorithms for topic discovery. This approach is beneficial because it limits the use of a labelled dataset for training purposes, which is labour intensive, time consuming, and expensive to obtain. There are popular topic modelling approaches, namely Latent Semantic Indexing (LSI), Probabilistic Latent Semantic Indexing (pLSI), and Latent Dirichlet Allocation (LDA). These models, especially LDA, have shown good performances over the years and several variants have been developed. Examples include the correlated topic model [14] and hierarchical Dirichlet process model [15]. However, these traditional TM approaches have been criticised for not being able to handle data sparsity, which is especially prevalent in short text. The Bi-Term Model (BTM) was developed to address this problem. This model has received praise in terms of performance when applied to short text. This is because BTM captures biterms from the whole corpus and, thus, the global co-occurrences of words are captured. The model assumes a pair of words can reveal topics much better than a single word occurrence. For example, biterms such as ATM_card and ATM_network are compared to unigrams such as card and network. However, Zhen et al. [16] stated BTM does not consider the comprehensive semantic dependencies of words, resulting in no contextual semantic. BTM has been criticised for the strong assumption made (that two co-occurring words will be assigned the same topic label) and this limits the performance. Thus, variants of this model have also been proposed such as Twitter-BTM [17] and GraphBTM [18]. More recently, the use of transformer-based language models has been proposed for TM. For example, BERT—Bidirectional Encoder Representations from Transformer [19]. However, these approaches are complex to use.

Despite the use of topic modelling becoming more popular, the examples in the literature found to have compared the TM techniques are very limited. In addition, there is no study found to have applied TM techniques in the Nigerian banking context. The banking industry is an important sector for every nation's economy and banking is considered a daily activity in the society [20]. The Nigerian banking sector heavily relies on the cash-based economy and most buying and selling are performed with physical cash [21]. Thus, the banks can benefit from topic modelling by helping them monitor their customers' preferences towards their product and service. In addition, this will help them understand what their customers are talking about. Although traditional approaches have shown good performance, their performance is inconsistent. Thus, there is a need to conduct further experimental comparative studies of TM approaches to validate state-of-the-art TM models. To this end, this study aims to compare topic modelling techniques for topic extraction in the Nigeria banking context. The rest of the paper is organised as follows. Section 2 will review the literature to provide background knowledge to this study. Section 3 will present the methodology. Section 4 will present and discuss the results and Section 5 will provide conclusions and recommendations.

## 2. Related Work

Topic models (TM) are considered more appropriate to extract topics because of their ability to extract implicit and explicit coherent topics. TM can be dated back to 1990 when Deerwester et al. [22] proposed Latent Semantic Indexing (LSI). LSI uses a singular value decomposition (SVD) of a large term–document matrix to identify a linear subspace such that the relationship between the term and document are captured. LSI was then adopted by various studies, such as Dewangan et al. [23]. However, LSI is limited as the technique does not assign a probability to the topic. In respect to this, Hofmann [24] developed probabilistic latent semantic indexing (pLSI). pLSI was also criticised for not being able to account for generative probabilistic models of the documents. Thus, Latent Dirichlet Allocation (LDA) was developed by Blei et al. [2] as an improvement.

LDA is a generative probabilistic model that captures the important intra-word/document statistical structure via a mixing distribution. LDA assumes that each document is associated with a topic distribution. Thus, topic assignment strongly relies on local co-occurrence. LDA has been used extensively for topic modelling [25–29]. For example, Çalli and Çalli [30] applied LDA to a dataset containing 10,594 airline-customers complaints from two Turkish airlines during the COVID-19 pandemic. Their study generated seven topics as the latent topics that customers complaints were about and used a qualitative human interpretation approach to validate the topics. Several studies [31–34] employed LDA to extract topics from a financial policy statements document. Moro et al. [35] utilised LDA to discover topics from 219 business intelligence (BI) articles within the banking domain and found that the most prominent topic in the BI banking literature is credit. Westerlund et al. [36] used LDA to generate six topics from 2702 comments of signers of the e-petition against the Bank of America. Tabiaa and Madani [37] utilised LDA to extract topics from eight Morocco mobile banking app reviews and showed security, services, quality, and interface as the topics customers talked about. Damane [38] applied LDA to generate topics from 26 monetary policy statements of Lesotho's central bank. Bastani et al. [39] utilised LDA to extract 40 topics from a dataset containing 86,803 consumer financial protection Bureau (CFPB) consumer complaints. This is beneficial to CFPB to understand and monitor what aspect of financial service the complaint narrative was about. However, the authors did not evaluate how well LDA has performed in their experimentation. Additionally, they have used trial and error to determine the number of K topics that does not efficiently demonstrate the optimal number of K topics. It is worth mentioning that the literature review findings showed that some of the studies found determined their number of topics through trial and error during experimentation. Gan and Qi [40] evidenced the need to select an optimal number of topics as this enhances predicting ability, high isolation within topics, repeatability, and no duplicated topics. Thus, an efficient approach to determining the optimal number of topics should be adopted.

More specifically, Hristova [41] extracted topics from Bulgarian bank chat data collected between January 2019 and April 2020. The author utilised cleaned 12,439 chats represented by a term frequency and inverse document frequency (TF-IDF) matrix to fit LDA. The study produced six coherent topics with a coherence score of 0.66. The result produced was used to understand the main themes that customers discussed, and the themes were profiled as loans, general information, digital banking, currency operations, identification, and cards/transfer. Despite the success achieved by LDA over the years, the unsupervised model has been criticized as it generates topics that contain irrelevant features and noisy topics [42], especially when dealing with short text such as social media text. Thus, LDA had been reported to suffer from the data sparsity problem. Improved extensions of LDA have been proposed such as the correlated topic model [14], the hierarchical Dirichlet process model [43], constrained-LDA [44], MaxEnt-LDA [45], automated knowledge LDA [46], LDADE [12], and ontology-LDA [42]. To overcome the data sparsity problem, Yan et al. [47] proposed the Bi-term Topic Model (BTM) for topic modelling, specifically for short text. BTM learns topics by modelling word pairs in the whole corpus. For example, "*my sure bank*" have biterms such as "*my sure, sure bank, my bank*". This means global occurrences of biterms are captured. BTM was developed with the assumption that two words will be assigned the same topic label if they have co-occurred [48]. However, it is worth mentioning that this study considers that assumption too strong. For example, a tweet such as "*My atm card could not work with bank X machine, not sure if it's the atm network or card issue*". In the tweet, the key words are "*atm, network, machine, card, issue*" and can have equal global co-occurrence. Unfortunately, topic modelling of tweets such as this suggest that the assumption may be too strong and does not apply in all cases. BTM has been shown to outperform traditional topic models such as LDA. For example, Discriminative-BTM [48], Twitter-BTM [17], and GraphBTM [18].

Due to the instability in the performance of the conventional TM approaches, the use of language models is evolving. Bidirectional Encoder Representations from Transformers (BERT) is a deep bidirectional unsupervised language representation model developed by Google [19] and has shown good result in topic extraction. For example, Yanuar and Shiramatsu [49] employed BERT for aspect extraction. They collected Indonesian tourism reviews from TripAdvisor and pretrained BERT with 4220 review sentences (using train batch size 32, max sequence length 128, number of epoch 16 learning rate $3 \times 10^{-5}$, and Adam epsilon $3 \times 10^{-8}$). Their approach used 501 Indonesian amusement park reviews for testing and reported an accuracy of 79.9% and F1 score of 73.8%. Bensoltane and Zaki [50] compared variants of BERT against bidirectional long short-term memory (BiLSTM) and conditional random field (CRF) for aspect extraction using 2265 Arabic news posts (retrieved from Facebook about the 2014 Gaza attack). They showed that the combination of BERT, bidirectional gated recurrent unit (BiGRU), and CRF achieved the highest performance for aspect term extraction with an F1 score (88%). However, despite the good performances shown by BERT, Liu et al. [51] stated BERT was undertrained and, thus, proposed RoBERTa, a robustly optimized BERT pretraining approach for topic extraction. They showed RoBERTa performed better than BERT on GLUE and SQuAD. Zhu et al. [52] inserted topic layers into fine-tuned RoBERTa and, thus, proposed a topic augmented language model for topic extraction. Grootendorst [53] presented BERTopic by leveraging clustering techniques and a class-based variation of TF-IDF to generate coherent topic representations. The author created document embeddings using a pretrained transformer-based language model to obtain document-level information. Thereafter, they applied dimension reduction in document embeddings and created semantically similar document clusters. Finally, the class-based version of TF-IDF was used to extract the topic representation from each topic. Grootendorst [53] showed the approach produced coherent topics in three different datasets namely, 20Newsgroups (16,309 news articles across 20 categories), BBC News (contains 2225 documents between 2004 and 2005), and Trump's tweets. Abuzayed and Al-Khalifa [54] compared BERTopic, LDA, and Non-Negative Matrix Factorization (NMF) in three different Arabic newspapers, namely, Assabah, Hespress, and Akhbarona. They showed BERTopic outperformed other models. Silveira et al. [13] successfully applied BERTopic to perform aspect extraction by investigating the stochastic topic modelling approaches for legal documents. Raju et al. [55] compared LDA, LSA, and BERTopic using the consumer financial protection Bureau dataset. They showed BERTopic outperformed other models with a coherence score of 0.33.

Despite the various approaches to topic modelling. The literature reviewed shows that there are limited studies on TM in the banking context. Secondly, there are no examples in the literature found to have applied or compared these TM techniques in the Nigerian banking context. The work of Hristova [41] is the closest research work to this study. However, it differs with the aim and data utilised because the study applied LDA to generate topics from Bulgarian bank chat data.

## 3. Methodology

We propose the use of Kernel Principal Component Analysis (KernelPCA) and K-means clustering in the BERTopic architecture for topic modelling. We conducted an experimental comparison of our approach to other topic models such as Latent Dirichlet Allocation (LDA), Latent Semantic Indexing (LSI), Hierarchical Dirichlet Process (HDP), and the variations of BERTopic. This will help to identify how the traditional TM techniques differ to the language topic models that utilise word/document embeddings.

### 3.1. Kernel Principal Component Analysis

The principal components of input data X can be obtained by solving the eigenvalue problem of the covariance matrix. With a sample of n observations, $x_i = x_1, x_2 \ldots x_n$. $x_i \in R^N$. Given that PCA operates on centred data, that is $\sum_{i=1}^{N} x_i = 0$, PCA diagonalizes the covariance matrix C as:

$$C = \frac{1}{N} \sum_{i=1}^{N} x_i x_i^T \quad (1)$$

The principal components are then obtained by solving eigenvalue problem:

$$Cv = \lambda v \quad (2)$$

In cases where n cannot be linearly separable, PCA is performed on a dot product space F (feature space) instead of the input space $R^N$ by mapping from $R^N$ to F. This can be illustrated as $\varphi: R^N \to F$ where $\varphi$ is the feature mapping of the input data to a high dimensional mapping. Then, $\sum_{i=1}^{N} \varphi(x_i) = 0$ PCA; the covariance matrix C in feature space F can be written as:

$$C = \frac{1}{N} \sum_{i=1}^{N} \varphi(x_i) \varphi(x_i)^T \quad (3)$$

The principal components are then obtained by solving the eigenvalue problem in Equation (2) above, where eigenvector v is spanned in F as:

$$v = \sum_{i=1}^{N} \alpha_i \phi(x_i) \quad (4)$$

Using the inner product between the two points, we can create the kernel:

$$K(x_i, x_j) = \phi(x_i)^T \phi(x_j) \quad (5)$$

where $K_{ij} = K(x_i, x_j)$ and $\alpha = (\alpha_1 \ldots \ldots \alpha_N)^T$. For the kernel component extraction, we compute the projection of each data sample x onto eigenvector V:

$$(\phi(x), V) = \sum_{i=1}^{N} \alpha_i K(x_i, x) \quad (6)$$

For any eigenvalue of C, $\lambda \geq 0$, the formulation of the eigenvalue problem for kernel matrix k can thus be defined as:

$$N\lambda\alpha = k\alpha \quad (7)$$

### 3.2. K-Means Clustering

K-means clustering is a popular clustering algorithm that discovers patterns within objects using their similar attributes. The algorithm requires the knowledge of K number of clusters to be predefined and the initial seed points are of importance. K-means clustering aims to partition n observation into K clusters in which each observation belongs to the cluster with the nearest means. Given X is a set of observations that contain n observation ($x_1, x_2, \ldots \ldots x_n$) in P dimensional vector, the algorithm follows the following steps:

- Declare K initial seed points (initial centroids) defined in P-dimensional vectors ($s_1, s_2\ s_p$) and the squared Euclidean distance between the ith object and the kth seed vector is obtained:

$$d^2_{(i,k)} = \sum_{j=1}^{P} \left(x_{ij} - s_j^{(k)}\right)^2 \quad (8)$$

- Cluster centroids are assigned and all objects are assigned to the closest centroids.
- The centroids changes to move closer to the centre point and the objects are reassigned to the closest centroid.
- Repeat the last 2 steps until no objects can be moved between clusters.

### 3.3. Dataset

We employed the data collected in the study of Ogunleye [56], which is publicly accessible in the Kaggle repository (https://www.kaggle.com/datasets/batoog/bank-customer-tweets-10000). Ogunleye [56] collected a total of nine hundred and fifty-nine thousand (959,000) Nigerian bank customer tweets for a duration of nine months (from 12 May 2019 to February 2020). We utilised 10,000 randomly sampled bank customers' tweets for the purpose of this study. This was performed because we considered having distinct tweets and avoided retweets to increase topic coverage. The texts are bank customer tweets towards the handles of 18 commercial banks in Nigeria. The tweets were in Pidgin English and English language. The dataset is particularly challenging not only because of the shortness of the text but the difference in phraseology. Pidgin English is an unofficial language widely used across West African countries such as Nigeria, Ghana, Cameroun, Equatorial Guinea, and Sierra Leone. In Nigeria, Pidgin English is the second most used language across all tribes after English. This is because Nigeria has over 500 different languages. Therefore, Pidgin was adopted as a common language across the tribes. The text (tweet) data were cleaned and pre-processed in Python. Python was adopted due to its prevalence in data science. The programming language is syntactically simple, simple to learn, and straightforward. Additionally, Python is open source, free to use, and has a rich ecosystem of libraries for scientific computing.

*3.4. Experimental Setup*

The dataset went through pre-processing. The natural language toolkit (NLTK) library [57] was used for tokenisation, removal of stop-words, word lemmatisation, and part of speech (POS) tagging. The POS were extracted using n-gram for n = 1, 2, 3. This means the unigram, bigram, and trigram were all considered. It is worth noting that steps such as word lemmatisation and POS tagging were applied to the English tokens in the dataset. The topic models were iterated to classify 10 topics and 20 terms per topic. The Gensim package [58] was used to implement the traditional topic models. The LDA model chunksize was set to 1740, with iterations at 1000 and passes at 20. Both alpha and beta Dirichlet priors were set to 'auto' to allow the model to automatically learn the best values for the hyperparameters during training. The LSI model chunksize was set to 1740 and the power iteration at 1000. The HDP model chunksize was set to 256, alpha at 1, and eta at 0.01. The BERTopic architecture [53] comprises the embedding, dimensionality reduction, clustering, and the C-TFIDF modules. In the embedding component, the BERT, SBERT [59], and FinBERT [60] were used to create word/document embedding. Principal Component Analysis (PCA), Kernel Principal Component Analysis (KPCA), Isometric Mapping (ISOMAP), Singular Value Decomposition (SVD), and Uniform Manifold Approximation and Projection (UMAP) were used for dimensionality reduction. Table 1 below shows the parameter set up of the dimensionality reduction algorithms.

**Table 1.** Dimensionality reduction algorithms.

| Algorithm | Experimental Set Up |
| --- | --- |
| ISOMAP | number of neighbours = 15, n_components = 5, eigen_solver = 'auto' |
| PCA | n_components = 5 |
| KPCA | n_components = 5, kernel = 'radial basis function', gamma = 15, random_state = 42 |
| SVD | n_components = 5, algorithm = 'randomized', random_state = 0 |
| UMAP | number of neighbours = 15, n_components = 10, min_dist = 0.0, metric = 'cosine' |

For the third component, the clustering methods used were K-means, Spectral, Agglomerative, MeanShift, Hierarchical Density-Based Spatial Clustering of Applications with Noise (HDBSCAN), Density-Based Spatial Clustering of Applications with Noise (DBSCAN), Balanced Iterative Reducing and Clustering using Hierarchies (BIRCH), and Ordering Points to Identify the Clustering Structure (OPTICS). Table 2 below shows the parameter set up of the cluster analysis algorithms.

For evaluation purposes, there is no agreed standard metric; however, the coherence score is used to justify the performance of the topic models. Topic coherence is a common evaluation metric for TM techniques and it is understood that the higher the coherence scores the better [61]. Röder et al. [62] showed that the coherence score has the highest correlation with human judgements of topics. However, Asghari et al. [63] argued that evaluating TM methods using a coherence score only is not enough. This needs to be complemented with topic quality and interpretability. In addition, there is no standard performance cut off for the coherence score to determine the best performing topic model. In the next section, we present the results of the experiment.

**Table 2.** Clustering algorithms.

| Algorithm | Experimental Set Up |
| --- | --- |
| K-means Clustering | n_cluster = 8, n_init = 10, max_iter = 300, init = 'k-means++' |
| Agglomerative Clustering | n_cluster = 2, linkage = 'ward', metric = 'Euclidean' |
| HDBSCAN | min_cluster_size = 10, metric = 'Euclidean', cluster_selection_method = 'eom', alpha = 1.0 |
| DBSCAN | Eps = 0.30, min_samples = 9, metric = 'Euclidean' |
| BIRCH | Threshold = 0.01, n_cluster = 3 |
| OPTICS | min_samples = 10, Eps = 0.8, metric = 'minkowski' |
| Meanshift | max_iter = 300, min_bin_freq = 1 |
| Spectral Clustering | n_cluster = 8, eigen_solver = 'arpack', assign_labels = 'kmeans', n_init = 10 |

## 4. Results

Table 3 below presents the coherence scores of the topic modelling techniques. The LDA has a coherence score of 0.3919 (cv),−9.1174 (umass), and produced interpretable terms as shown in Table 4 below.

**Table 3.** Model coherence score.

| Model | Coherence Score (Cv) | Coherence Score (Umass) |
| --- | --- | --- |
| LDA | 0.3919 | −9.1174 |
| LSI | 0.3912 | −2.6630 |
| HDP | 0.6347 | −16.5378 |

**Table 4.** LDA Topics.

| LDA | Terms |
| --- | --- |
| Topic 1 | customer, cenbank_central, branch, time, month, people, guy, atm, even, week, still, day, service, call, complaint, nothing, customer_care, money, customer_service, thing |
| Topic 2 | job, morgan, free, min, company, una, fossil_fuel, ready, hey_investment, gas_coal, financing_oil, earth_stop, company_nofossil, nofossilfuel, fargo_jpmorgan, change, actually, head, woman, country |
| Topic 3 | help, access, account, money, transaction, please, transfer, pls, yet, day, number, kindly, yesterday, back, today, issue, alert, need, airtime, fund |
| Topic 4 | loan, help, veilleur, bbnaija, cenbank_central, kindly_follow, day, send_direct, officialefcc, message_details, monthly, thank, house, enable_assist, salary, ph, advise_appropriately, heritagebank, eyin_ole, force |
| Topic 5 | card, charge, never, new, use, open, year, debit, later, small, fee, pay, dey, top, see, right, also, month, nyou, far |
| Topic 6 | business, lagos, great, school, news, monetizable_activity, world, additional_information, soon, let, part, good, best, opportunity, today, deposit, wish, thanks, trump, event |
| Topic 7 | dm, help, detail, thank, number, twitter, kindly, please, complaint, sorry, send, tweet, transaction, handle, account, information, send_dm, channel, phone_number, message |
| Topic 8 | line, phone, support, ng, mobile, mtnng_mtn, hope, food, consumer, government, international, buhari, sort, whatsapp, explanation, enable, lagos, foundation, industry, ubaat |
| Topic 9 | always, poor, stop, financial, ng_union, cool, youfirst, sport, business, service, finance, internet_banking, premiercool, mind, tell, thing, daily, single, fast, vista_intl |
| Topic 10 | credit_card, citi, bill, state, debt, payment, chase, omo_iya, mastercard, nthank, ng, future, true, alat, abuja, pay, india, atm_machine, child, product |

The above result showed HDP has the highest coherence score (cv), by far, of 0.6347. However, HDP achieved the highest negative coherence score (umass) of −16.5378. The terms were investigated manually as shown in Table 5 below. The HDP terms contain some unwanted words such as '*la*', '*aw*', and '*v*' despite the corpus filtering, cleaning, and pruning. In addition, we observed some terms were common amongst the topics. For example, the term '*help*' appeared in eight different topics. Whilst LSI performed with a marginally lower coherence score (cv) with low quality topics as shown in Table 6 below. This is because the LSI terms are repeated across the topics as shown in Table 6. For example, terms such as '*transaction*' and '*access*' appeared in at least three different topics, '*help*' appeared in five different topics, and '*money*' and '*account*' appeared in six different topics. LSI struggled with the distribution of terms. The model produced overlapping semantic terms across the different topics because LSI tried to map the relationship between the term and document to detect contextual semantic terms. In summary, the LDA produced terms that are more interpretable. Thus, we tune the number of topics to determine the number of topics with an optimal solution.

**Table 5.** HDP Topics.

| HDP | Terms |
| --- | --- |
| Topic 1 | help, access, account, money, transaction, please, dm, number, day, guy, still, app, la, time, kindly, director, blame, transfer, banking, yet |
| Topic 2 | oct, promise, still, baba, dont, thread, access, longer, v, gift, swift, start, fccpnigeria, delay, sent, reminder, partnership, mind, list, help |
| Topic 3 | facebook, association, knowledge, deduction, nhere, aware, help, complete, travel, away, assistance, delivery, goalcomnigeria, review, nation, fire, aw, fintech, right, peace |
| Topic 4 | product, notice, wonder, unclemohamz, operation, banker, exchange, cenbank_central, plz, otherwise, see, respectively, better, blue, internet, help, tired, customer_service, pain, deal |
| Topic 5 | black, blame, joke, help, branch, personally, negative, notice, merger, tuesday, okay, today, aw, everyday, market, rectify, private, executive, receiver, appropriately |
| Topic 6 | help, duty, angry, samuel, die, fintech, sorry_experience, worst, wizkidayo_wizkidayo, opay, nysc, gov, weekly, brilafm, glad, star, closing, ubafc, question, omo_iya |
| Topic 7 | true, space, fast, head, explorer, star, v, help, pastor, meant, etc, place, capital, still, sept, mbuhari_muhammad, aisha, war, sincerely, morning |
| Topic 8 | account, apology, olu, fact, tomorrow, news, power, noodle, advance, help, completely, document, pick, lately, mistake, mastercard, okay, life, lie, several |
| Topic 9 | hey_investment, experience, help, delete, peter, allowance, told, poor, access, reflect, unilorin, brilafm, complaint_enquiry, wallet, deduct, chidi, directly, imagine, enable_assist, capitalone |
| Topic 10 | form, resolution, report, concerned, cantstopwont, nplease, anyone, sister, offer, worth, awareness, october, keep, day, till_date, last, problem, ibrahim, shit, reno |

To achieve an optimal solution, the LDA was tuned by changing the parameter for the number of topics to be in the range from 5 to 250. The LDA as shown in Figure 1 below produced a coherence score at a close range from 0.3 to 0.65.

The BERTopic architecture was also employed, and the previously stated components were deployed during the experimentation. Table 7 below shows the approaches adopted in the BERTopic model and thus presents their result.

**Table 6.** LSI Topics.

| LSI | Terms |
|---|---|
| Topic 1 | help, access, account, money, transaction, please, number, day, dm, transfer, yet, guy, cenbank_central, kindly, pls, time, card, issue, still, back |
| Topic 2 | account, access, money, help, please, number, transaction, transfer, yet, kindly, back, pls, yesterday, guy, cenbank_central, naira, credit, detail, card, debit |
| Topic 3 | money, account, help, number, access, back, day, guy, cenbank_central, dm, refund, detail, people, still, yet, time, please, month, atm, reverse |
| Topic 4 | access, help, account, transaction, please, money, dm, kindly, number, thank, detail, complaint, diamond, pls, day, transfer, cenbank_central, twitter, central, response |
| Topic 5 | transaction, help, money, please, account, card, day, cenbank_central, yet, kindly, time, dm, detail, still, successful, debit, access, today, online, number |
| Topic 6 | please, transaction, pls, card, number, day, dm, help, account, transfer, money, time, access, thank, guy, kindly, even, detail, cenbank_central, people |
| Topic 7 | day, transaction, card, time, guy, money, cenbank_central, branch, people, still, customer, issue, month, even, debit, today, complaint, service, please, year |
| Topic 8 | card, day, debit, cenbank_central, please, guy, dm, people, branch, pls, month, time, transfer, use, charge, number, atm, yet, eyin_ole, ur_ur |
| Topic 9 | dm, number, day, card, guy, people, issue, time, nplease, please, detail, debit, kindly, transfer, help, check, customer, pls, even, account |
| Topic 10 | today, refund, people, news, day, business, transaction, number, citizen, transfer, global, pls, need, customer, citi, kindly, guy, month, america, please |

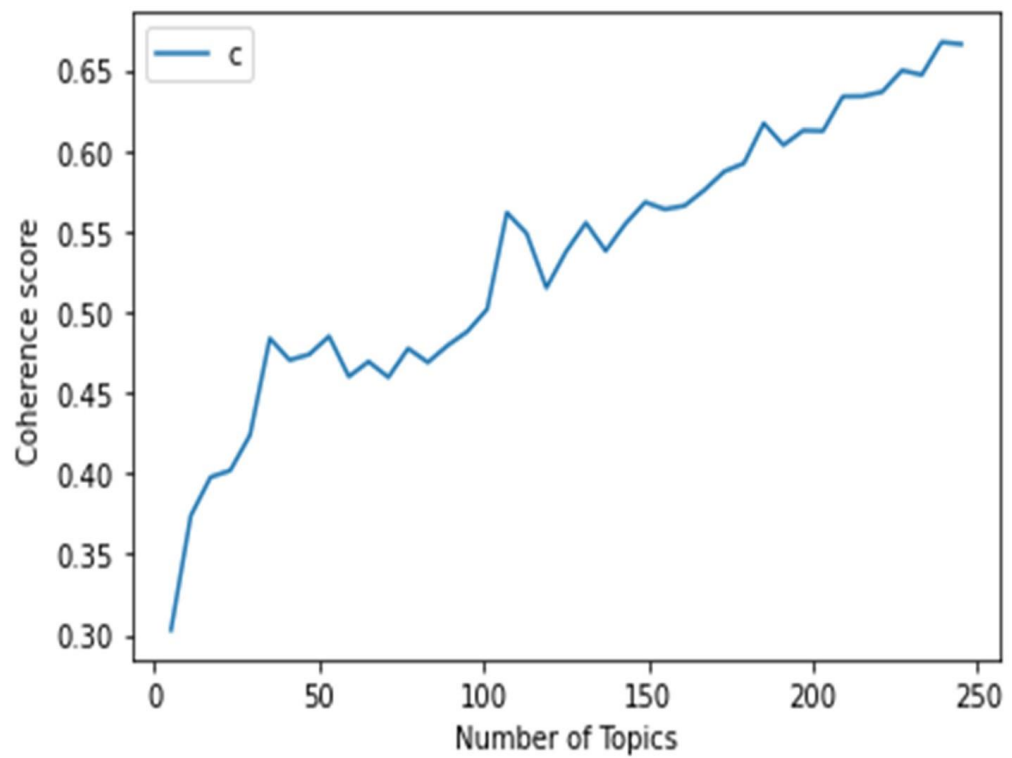

**Figure 1.** Plot of topics versus coherence score.

Table 7. BERTopic coherence score.

| Model | Component 1 (Embedding) | Component 2 (Dimension Reduction) | Component 3 (Clustering) | Coherence Score |
|---|---|---|---|---|
| BERTopic | BERT | Kernel PCA | BIRCH | 0.7071 |
| BERTopic | BERT | Kernel PCA | DBSCAN | 0.7528 |
| BERTopic | BERT | Kernel PCA | HDBSCAN | 0.4294 |
| BERTopic | BERT | Kernel PCA | Agglomerative Clustering | 0.4165 |
| BERTopic | BERT | Kernel PCA | OPTICS | 0.4902 |
| BERTopic | BERT | Kernel PCA | Spectral Clustering | 0.8451 |
| BERTopic | **BERT** | **Kernel PCA** | **K-means** | **0.8463** |
| BERTopic | BERT | ISOMAP | K-means | 0.4951 |
| BERTopic | BERT | PCA | K-means | 0.5251 |
| BERTopic | BERT | SVD | K-means | 0.5283 |
| BERTopic | BERT | UMAP | K-means | 0.5340 |
| BERTopic | BERT | UMAP | Spectral Clustering | 0.6267 |
| BERTopic | BERT | UMAP | MeanShift | 0.7522 |
| BERTopic | BERT | UMAP | HDBSCAN | 0.6705 |
| BERTopic | FinBERT | UMAP | HDBSCAN | 0.6667 |
| BERTopic | SBERT | UMAP | HDBSCAN | 0.6678 |

The experimental comparative result shows that the BERTopic with components BERT for embeddings, KernelPCA for dimension reduction, and K-means clustering achieved the highest coherence score of 0.8463. The kernel function can deal with non-linear data. The kernel method helps improve non-linearly separable data by increasing the dimension to a higher dimensional space such that the data can be linearly separable [64]. This kernel trick has been applied in the PCA to help produce five principal components that were fed into the partitioned clusters. In addition, the approach produced coherent terms as shown in Figure 2 below. The topics produced were around USSD code, general enquiry about account, transaction problem, mobile app, and ATMs. Some of the topics generated are very similar to those produced in the study of Hristova [41]. For example, general information, digital banking, and cards/transfer. It is worth noting that during the experimentation, the FinBERT was utilised to generate the document embeddings. This is because FinBERT was trained with financial data. Unfortunately, the FinBERT embeddings did not improve the result. Similarly, the SBERT embeddings did not improve the result.

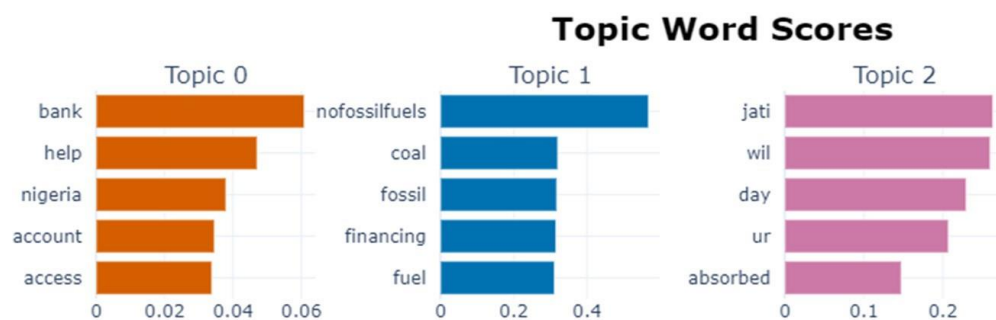

**Figure 2.** BERTopic terms.

In addition, it is worth noting that the number of topics were not specified for the results shown in Table 7 above. Thus, the number of topics produced were above 100. In practice, this is not usually the case, as it is not economical for service industries such as the banks to have over 100 unit of topics to be monitored. Based on that assumption, this study inputted 10 as the number of topics for the BERTopic. The result produced showed a significant drop in the coherence score. For example, BERTopic (BERT, UMAP, HDBSCAN, topic = 10, and terms = 20) produced 0.54, which is lower than 0.67 when topics and terms are not specified. Specifically, BERTopic (BERT, kernelPCA, K-means, topic = 10, and terms = 20) produced 0.76 compared to 0.8463 when the topics and terms are not specified. This result is consistent with the study of Raju et al. [55], Silveira et al. [13], Hristova [41], Abuzayed and Al-Khalifa [54], and Grootendorst [53] that showed BERTopic outperformed other models. Most importantly, the combined approach of KernelPCA and K-means has been shown to outperform other combined dimensionality reduction and clustering techniques. For example, Lyu et al. [65] performed a comparative study for solar irradiance forecasting and showed the combination of KernelPCA and K-means clustering outperforms combined KernelPCA and spectral clustering, KernelPCA and BIRCH, and KernelPCA and agglomerative clustering.

## 5. Conclusions

Since labelled datasets are expensive, labour intensive, time consuming, and not readily available to utilise supervised learning algorithms in this context, we validate a totally unsupervised approach for extracting topics. This is vital for service industries to understand emerging topics in their customer discussions. The purpose of the current study was to determine the state-of-the-art topic modelling technique in the Nigerian banking context. To this end, we have compared TM approaches and ascertained the use of kernel principal component analysis and k-means clustering in the BERTopic architecture as the state-of-the-art topic modelling technique. Our experimental results showed the LDA-produced coherence score ranged from 0.3 to 0.65 with the number of topics being set to a range between 5 and 250. The BERTopic (BERT, kernelPCA, and k-means clustering) achieved a coherence score of 0.76 when the number of topics is set to 10 and the terms is set to 20 (against leaving the number of topics and terms unspecified where it was 0.8463). Thus, our approach provided a reasonably high performance even restricting the number of topics to a more manageable level. In summary, the use of BERTopic as a language model is very promising as the topics produced were interpretable and of high-quality terms. The LDA with a well processed corpus performed well too but not better. The traditional topic model struggled with the different subtleties in Pidgin English, which depends on the native language of the tweeter. Most notably, the HDP and LSI produced overlapping terms in this context. Thus, this study proposes the use of BERTopic and recommends the application of BERTopic in various domains to compare against the traditional approaches.

*Contributions, Limitation, and Future Work*

In this study, we have prepared a new dataset using tweets from Nigerian bank customers. This dataset will enhance the research in topic discovery in the Nigerian banking context. Secondly, we explored the use of a language model (BERT), which requires little or no pre-processing of text (input) data to create document embeddings. Lastly, we demonstrated the use of kernel principal component analysis and k-means clustering for extracting topics. This study provided a comprehensive explanation of the methods deployed. To the best of our knowledge, this is the first study to compare TM approaches in the Nigerian banking context. Our study is useful for service industries to discover emerging topics. Specifically, the banks can adopt our approach to monitor their customers' preferences towards their product and service. For example, our findings showed topics around USSD code, general enquiry, mobile app, and ATMs were the leading themes in the bank customer discussion.

A major limitation is that there are limited lexical resources publicly available to support the pre-processing of Pidgin terms. Most notably, there is a need to pre-train the language models with Pidgin English words/documents to provide opportunities for research in that area. In the future, there is need to explore the use of the kernel trick on spectral clustering as the model has a within dimensionality reduction component. This is because, in our experiment, the combination of KPCA and spectral clustering also yielded a competitive result. Another area to explore in the future is the use of domain knowledge to generate pre-defined labels such that BERTopic can be deployed as a semi-supervised learning algorithm to generate N coherent topics. In addition, there is a need to create a labelled set to use supervised learning algorithms for topic discovery. This will enhance the evaluation processes by comparing the model prediction against the ground truth.

**Author Contributions:** Conceptualization, B.O.; methodology, B.O.; software B.O.; validation, B.O.; formal analysis, B.O.; investigation, B.O.; resources B.O.; data curation, B.O.; writing—original draft preparation, B.O.; writing—review and editing, T.M., L.H., J.G. and T.B.; visualization, B.O.; supervision, T.M., L.H., J.G. and T.B.; project administration, T.M., L.H., J.G. and T.B. All authors have read and agreed to the published version of the manuscript.

**Funding:** This research received no external funding.

**Institutional Review Board Statement:** Not applicable.

**Informed Consent Statement:** Not applicable.

**Data Availability Statement:** The dataset used in this work is available on Kaggle repository https://www.kaggle.com/datasets/batoog/bank-customer-tweets-10000 and the code https://www.kaggle.com/code/batoog/bertopic.

**Conflicts of Interest:** The authors declare no conflict of interest.

## References

1. Meng, Y.; Zhang, Y.; Huang, J.; Zhang, Y.; Han, J. Topic discovery via latent space clustering of pretrained language model representations. In Proceedings of the ACM Web Conference 2022, Lyon, France, 25–29 April 2022; pp. 3143–3152.
2. Blei, D.M.; Ng, A.Y.; Jordan, M.I. Latent dirichlet allocation. *J. Mach. Learn. Res.* **2003**, *3*, 993–1022.
3. Dandala, B.; Joopudi, V.; Devarakonda, M. Adverse drug events detection in clinical notes by jointly modeling entities and relations using neural networks. *Drug Saf.* **2019**, *42*, 135–146. [CrossRef] [PubMed]
4. Kastrati, Z.; Arifaj, B.; Lubishtani, A.; Gashi, F.; Nishliu, E. Aspect-Based Opinion Mining of Students' Reviews on Online Courses. In Proceedings of the 2020 6th International Conference on Computing and Artificial Intelligence, Tianjin, China, 23–26 April 2020; pp. 510–514.
5. Ray, P.; Chakrabarti, A. A mixed approach of deep learning method and rule-based method to improve aspect level sentiment analysis. *Appl. Comput. Informatics* **2020**, *18*, 163–178. [CrossRef]
6. Pennacchiotti, M.; Gurumurthy, S. Investigating topic models for social media user recommendation. In Proceedings of the 20th International Conference Companion on World Wide Web, Hyderabad, India, 28 March–1 April 2011; pp. 101–102.
7. Wang, D.; Zhu, S.; Li, T.; Gong, Y. Multi-document summarization using sentence-based topic models. In Proceedings of the ACL-IJCNLP 2009 Conference Short Papers, Proceedings of the Joint Conference of the 47th Annual Meeting of the Association for Computational Linguistics and 4th International Joint Conference on Natural Language Processing of the AFNLP, Singapore, 2–7 August 2009; World Scientific: Singapore, 2009; pp. 297–300.
8. Tepper, N.; Hashavit, A.; Barnea, M.; Ronen, I.; Leiba, L. Collabot: Personalized group chat summarization. In Proceedings of the Eleventh ACM International Conference on Web Search and Data Mining, Los Angeles, CA, USA, 5–9 February 2018; pp. 771–774.
9. Sabeeh, V.; Zohdy, M.; Bashaireh, R.A. Fake News Detection Through Topic Modeling and Optimized Deep Learning with Multi-Domain Knowledge Sources. In *Advances in Data Science and Information Engineering*; Springer: Cham, Switzerland, 2021; pp. 895–907.
10. Wang, T.; Huang, Z.; Gan, C. On mining latent topics from healthcare chat logs. *J. Biomed. Inform.* **2016**, *61*, 247–259. [CrossRef]
11. Adanir, G.A. Detecting topics of chat discussions in a computer supported collaborative learning (CSCL) environment. *Turk. Online J. Distance Educ.* **2019**, *20*, 96–114. [CrossRef]
12. Agrawal, A.; Fu, W.; Menzies, T. What is wrong with topic modeling? And how to fix it using search-based software engineering. *Inf. Softw. Technol.* **2018**, *98*, 74–88. [CrossRef]
13. Silveira, R.; Fernandes, C.G.; Neto, J.A.M.; Furtado, V.; Pimentel Filho, J.E. Topic modelling of legal documents via legal-bert. In Proceedings of the CEUR Workshop, Virtual Event, College Station, TX, USA, 19–20 August 2021; ISSN 1613-0073. Available online: http://ceur-ws.org (accessed on 12 October 2022).
14. Blei, D.M.; Lafferty, J.D. A correlated topic model of science. *Ann. Appl. Stat.* **2007**, *1*, 17–35. [CrossRef]
15. Teh, Y.W.; Jordan, M.I.; Beal, M.J.; Blei, D.M. Hierarchical dirichlet processes. *J. Am. Stat. Assoc.* **2006**, *101*, 1566–1581. [CrossRef]
16. Zhen, L.; Yabin, S.; Ning, Y. A Short Text Topic Model Based on Semantics and Word Expansion. In Proceedings of the 2022 IEEE 2nd International Conference on Computer Communication and Artificial Intelligence (CCAI), Beijing, China, 6–8 May 2022; pp. 60–64.
17. Chen, W.; Wang, J.; Zhang, Y.; Yan, H.; Li, X. User based aggregation for biterm topic model. In Proceedings of the 53rd Annual Meeting of the Association for Computational Linguistics and the 7th International Joint Conference on Natural Language Processing, Beijing, China, 26–31 July 2015; Volume 2, pp. 489–494.

18. Zhu, Q.; Feng, Z.; Li, X. GraphBTM: Graph enhanced autoencoded variational inference for biterm topic model. In Proceedings of the Conference on Empirical Methods in Natural Language Processing (EMNLP), Brussels, Belgium, 31 October–4 November 2018.
19. Devlin, J.; Chang, M.W.; Lee, K.; Toutanova, K. BERT: Pre-Training of Deep Bidirectional Transformers for Language Understanding. *arXiv* **2018**, arXiv:1810.04805.
20. Alsmadi, A.A.; Sha'Ban, M.; Al-Ibbini, O.A. The Relationship between E-Banking Services and Bank Profit in Jordan for the Period of 2010–2015. In Proceedings of the 2019 5th International Conference on E-Business and Applications, Bangkok, Thailand, 25–28 February 2019; pp. 70–74.
21. Ailemen, I.O.; Enobong, A.; Osuma, G.O.; Evbuomwan, G.; Ndigwe, C. Electronic banking and cashless policy in Nigeria. *Int. J. Civ. Eng. Technol.* **2018**, *9*, 718–731.
22. Deerwester, S.; Dumais, S.T.; Furnas, G.W.; Landauer, T.K.; Harshman, R. Indexing by latent semantic analysis. *J. Am. Soc. Inf. Sci.* **1990**, *41*, 391–407. [CrossRef]
23. Dewangan, J.K.; Sharaff, A.; Pandey, S. Improving topic coherence using parsimonious language model and latent semantic indexing. In *ICDSMLA 2019*; Springer: Singapore, 2020; pp. 823–830.
24. Hofmann, T. Probabilistic latent semantic indexing. In Proceedings of the 22nd Annual International ACM SIGIR Conference on Research and Development in Information Retrieval, Berkeley, CA, USA, 15–19 August 1999; pp. 50–57.
25. Alfieri, L.; Gabrielyan, D. *The Communication Reaction Function of the European Central Bank. An Analysis Using Topic Modelling*; Eesti Pank: Tallinn, Estonia, 2021.
26. Bertalan, V.G.; Ruiz, E.E.S. Using topic modeling to find main discussion topics in Brazilian political websites. In Proceedings of the 25th Brazilian Symposium on Multimedia and the Web, Rio de Janeiro, Brazil, 29 October–1 November 2019; pp. 245–248.
27. Blei, D.M. Probabilistic topic models. *Commun. ACM* **2012**, *55*, 77–84. [CrossRef]
28. Kastrati, Z.; Kurti, A.; Imran, A.S. WET: Word embedding-topic distribution vectors for MOOC video lectures dataset. *Data Brief* **2020**, *28*, 105090. [CrossRef] [PubMed]
29. Qi, B.; Costin, A.; Jia, M. A framework with efficient extraction and analysis of Twitter data for evaluating public opinions on transportation services. *Travel Behav. Soc.* **2020**, *21*, 10–23. [CrossRef]
30. Çallı, L.; Çallı, F. Understanding Airline Passengers during COVID-19 Outbreak to Improve Service Quality: Topic Modeling Approach to Complaints with Latent Dirichlet Allocation Algorithm. *Res. Rec. J. Transp. Res. Board* **2022**. [CrossRef]
31. Doh, T.; Kim, S.; Yang, S.-K.X. How You Say It Matters: Text Analysis of FOMC Statements Using Natural Language Processing. *Fed. Reserv. Bank Kans. City Econ. Rev.* **2021**, *106*, 25–40. [CrossRef]
32. Edison, H.; Carcel, H. Text data analysis using Latent Dirichlet Allocation: An application to FOMC transcripts. *Appl. Econ. Lett.* **2020**, *28*, 38–42. [CrossRef]
33. Lee, H.; Seo, H.; Geum, Y. Uncovering the topic landscape of product-service system research: From sustainability to value creation. *Sustainability* **2018**, *10*, 911. [CrossRef]
34. Shirota, Y.; Yano, Y.; Hashimoto, T.; Sakura, T. Monetary policy topic extraction by using LDA: Japanese monetary policy of the second ABE cabinet term. In Proceedings of the 2015 IIAI 4th International Congress on Advanced Applied Informatics, Okayama, Japan, 12–16 July 2015; pp. 8–13.
35. Moro, S.; Cortez, P.; Rita, P. Business intelligence in banking: A literature analysis from 2002 to 2013 using text mining and latent Dirichlet allocation. *Expert Syst. Appl.* **2015**, *42*, 1314–1324. [CrossRef]
36. Westerlund, M.; Olaneye, O.; Rajahonka, M.; Leminen, S. Topic modelling on e-petition data to understand service innovation resistance. In Proceedings of the International Society for Professional Innovation Management (ISPIM) Conference, Palazzo dei Congressi, Florence, Italy, 4–7 June 2019; pp. 1–13.
37. Tabiaa, M.; Madani, A. Analyzing the Voice of Customer through online user reviews using LDA: Case of Moroccan mobile banking applications. *Int. J. Adv. Trends Comput. Sci. Eng.* **2021**, *10*, 32–40. [CrossRef]
38. Damane, M. Topic Classification of Central Bank Monetary Policy Statements: Evidence from Latent Dirichlet Allocation in Lesotho. *Acta Univ. Sapientiae Econ. Bus.* **2022**, *10*, 199–227. [CrossRef]
39. Bastani, K.; Namavari, H.; Shaffer, J. Latent Dirichlet allocation (LDA) for topic modeling of the CFPB consumer complaints. *Expert Syst. Appl.* **2019**, *127*, 256–271. [CrossRef]
40. Gan, J.; Qi, Y. Selection of the Optimal Number of Topics for LDA Topic Model—Taking Patent Policy Analysis as an Example. *Entropy* **2021**, *23*, 1301. [CrossRef] [PubMed]
41. Hristova, G. Topic modelling of chat data: A case study in the banking domain. *AIP Conf. Proc.* **2021**, *2333*, 150014.
42. Ali, F.; Kwak, D.; Khan, P.; El-Sappagh, S.; Ali, A.; Ullah, S.; Kwak, K.S. Transportation sentiment analysis using word embedding and ontology-based topic modeling. *Knowl.-Based Syst.* **2019**, *174*, 27–42. [CrossRef]
43. Teh, Y.; Jordan, M.; Beal, M.; Blei, D. Sharing clusters among related groups: Hierarchical dirichlet processes. In *Advances in Neural Information Processing Systems 17, Proceedings of the Neural Information Processing Systems, NIPS 2004, Vancouver, BC, Canada, 13–18 December 2004*; ACM: New York, NY, USA, 2004.
44. Zhai, Z.; Liu, B.; Xu, H.; Jia, P. Constrained LDA for grouping product features in opinion mining. In *Pacific-Asia Conference on Knowledge Discovery and Data Mining*; Springer: Berlin/Heidelberg, Germany, 2011; pp. 448–459.
45. Zhao, X.; Jiang, J.; Yan, H.; Li, X. Jointly modeling aspects and opinions with a MaxEnt-LDA hybrid. In Proceedings of the Conference on Empirical Methods in Natural Language, Cambridge, MA, USA, 9–11 October 2010.

46. Chen, Z.; Mukherjee, A.; Liu, B. Aspect extraction with automated prior knowledge learning. In Proceedings of the 52nd Annual Meeting of the Association for Computational Linguistics, Baltimore, MD, USA, 22–27 June 2014; Volume 1, pp. 347–358.
47. Yan, X.; Guo, J.; Lan, Y.; Cheng, X. A biterm topic model for short texts. In Proceedings of the 22nd International Conference on World Wide Web, Rio de Janeiro, Brazil, 13–17 May 2013; pp. 1445–1456.
48. Xia, Y.; Tang, N.; Hussain, A.; Cambria, E. Discriminative bi-term topic model for headline-based social news clustering. In Proceedings of the Twenty-Eighth International Flairs Conference, Hollywood, FL, USA, 18–25 May 2015.
49. Yanuar, M.R.; Shiramatsu, S. Aspect extraction for tourist spot review in Indonesian language using BERT. In Proceedings of the 2020 International Conference on Artificial Intelligence in Information and Communication (ICAIIC), Fukuoka, Japan, 19–21 February 2020; pp. 298–302.
50. Bensoltane, R.; Zaki, T. Towards Arabic aspect-based sentiment analysis: A transfer learning-based approach. *Soc. Netw. Anal. Min.* **2022**, *12*, 7. [CrossRef]
51. Liu, Y.; Ott, M.; Goyal, N.; Du, J.; Joshi, M.; Chen, D.; Stoyanov, V. Roberta: A robustly optimized bert pretraining approach. *arXiv* **2019**, arXiv:1907.11692.
52. Zhu, L.; Pergola, G.; Gui, L.; Zhou, D.; He, Y. Topic-Driven and Knowledge-Aware Transformer for Dialogue Emotion Detection. *arXiv* **2021**, arXiv:2106.01071.
53. Grootendorst, M. BERTopic: Neural topic modelling with a class-based TF-IDF procedure. *arXiv* **2022**, arXiv:2203.05794.
54. Abuzayed, A.; Al-Khalifa, H. BERT for Arabic topic modeling: An experimental study on BERTopic technique. *Procedia Comput. Sci.* **2021**, *189*, 191–194. [CrossRef]
55. Raju, S.V.; Bolla, B.K.; Nayak, D.K.; Kh, J. Topic Modelling on Consumer Financial Protection Bureau Data: An Approach Using BERT Based Embeddings. In Proceedings of the 2022 IEEE 7th International Conference for Convergence in Technology (I2CT), Mumbai, India, 7–9 April 2022; pp. 1–6.
56. Ogunleye, B.O. Statistical Learning Approaches to Sentiment Analysis in the Nigerian Banking Context. Ph.D. Thesis, Sheffield Hallam University, Sheffield, UK, 2021.
57. Bird, S.; Klein, E.; Loper, E. *Natural Language Processing with Python: Analysing Text with the Natural Language Toolkit*; O'Reilly Media, Inc.: Sebastopol, CA, USA, 2009.
58. Rehurek, R.; Sojka, P. Software framework for topic modelling with large corpora. In Proceedings of the LREC 2010 Workshop on New Challenges for NLP Frameworks, Valletta, Malta, 22 May 2010.
59. Reimers, N.; Gurevych, I. Sentence-bert: Sentence embeddings using siamese bert-networks. *ArXiv* **2019**, arXiv:1908.10084.
60. Araci, D. FinBERT: Financial Sentiment Analysis with Pre-Trained Language Models. *arXiv* **2019**, arXiv:1908.10063.
61. Albalawi, R.; Yeap, T.H.; Benyoucef, M. Using Topic Modeling Methods for Short-Text Data: A Comparative Analysis. *Front. Artif. Intell.* **2020**, *3*, 42. [CrossRef] [PubMed]
62. Röder, M.; Both, A.; Hinneburg, A. Exploring the space of topic coherence measures. In Proceedings of the Eighth ACM International Conference on Web Search and Data Mining, Shanghai, China, 2–6 February 2015; pp. 399–408.
63. Asghari, M.; Sierra-Sosa, D.; Elmaghraby, A.S. A topic modeling framework for spatio-temporal information management. *Inf. Process. Manag.* **2020**, *57*, 102340. [CrossRef]
64. Schölkopf, B.; Smola, A.; Müller, K.R. Kernel principal component analysis. In *International Conference on Artificial Neural Networks*; Springer: Berlin/Heidelberg, Germany, 1997; pp. 583–588.
65. Lyu, C.; Basumallik, S.; Eftekharnejad, S.; Xu, C. A data-driven solar irradiance forecasting model with minimum data. In Proceedings of the 2021 IEEE Texas Power and Energy Conference (TPEC), College Station, TX, USA, 2–5 February 2021; pp. 1–6.